# U-Net and its variants for medical image segmentation: theory and applications

**Nahian Siddique [1], Paheding Sidike [2], Colin Elkin [1] and Vijay Devabhaktuni [1]**

[1] Department of Electrical and Computer Engineering, Purdue University Northwest
[2] Department of Applied Computing, Michigan Technological University

**Abstract:** U-net is an image segmentation technique developed primarily for medical image analysis that can precisely segment images using a scarce amount of training data. These traits provide U-net with a very high utility within the medical imaging community and have resulted in extensive adoption of U-net as the primary tool for segmentation tasks in medical imaging. The success of U-net is evident in its widespread use in all major image modalities from CT scans and MRI to X-rays and microscopy. Furthermore, while U-net is largely a segmentation tool, there have been instances of the use of U-net in other applications. As U-net's potential is still increasing, in this review we look at the various developments that have been made in the U-net architecture and provide observations on recent trends. We examine the various innovations that have been made in deep learning and discuss how these tools facilitate U-net. Furthermore, we look at image modalities and application areas where U-net has been applied.

## 1. Introduction

Thanks to recent advances in deep learning in computer vision within the last decade, deep learning has been increasingly utilized in the analysis of medical images. Whilst the use of deep learning in computer vision has seen rapid growth in many different fields, it still faced some challenges in the medical imaging field. There have been many breakthrough techniques over the years to overcome these various challenges. Many techniques and methods have been improvised to developed to such challenges. One such technique that will be discussed in this review will be the U-net, a deep learning technique widely adopted within the medical imaging community.

U-net is a neural network architecture designed primarily for image segmentation [1]. The basic structure of a U-net architecture consists of two paths. The first path is the contracting path, also known as the encoder or the analysis path, which is similar to a regular convolution network and provides classification information. The second is an expansion path, also known as the decoder or the synthesis path, consisting of up-convolutions and concatenations with features from the contracting path. This expansion allows the network to learn localized classification information. Additionally, the expansion path also increases the resolution of the output which can then be pass onto a final convolutional layer to create a fully segmented image. The resulting network is almost symmetrical, giving it a u-like shape. The main canonical task performed by most convolutional networks is to classify the whole image into a single label. However, classification networks fail to provide pixel-level context information which is much needed in medical image analysis. While there have been previous attempts at segmentation tasks, it wasn't until U-net by Ronneberger et al.[1] that there was a significant improvement in medical image segmentation performance. The U-net network was developed based on the works of Long, J et al. [2] using fully convolutional networks. Their implementation achieved better performance than the previous best on the ISBI 2012 challenge and won the ISBI cell tracking challenge in 2015, beating the state of the ray at the time with a considerable margin.

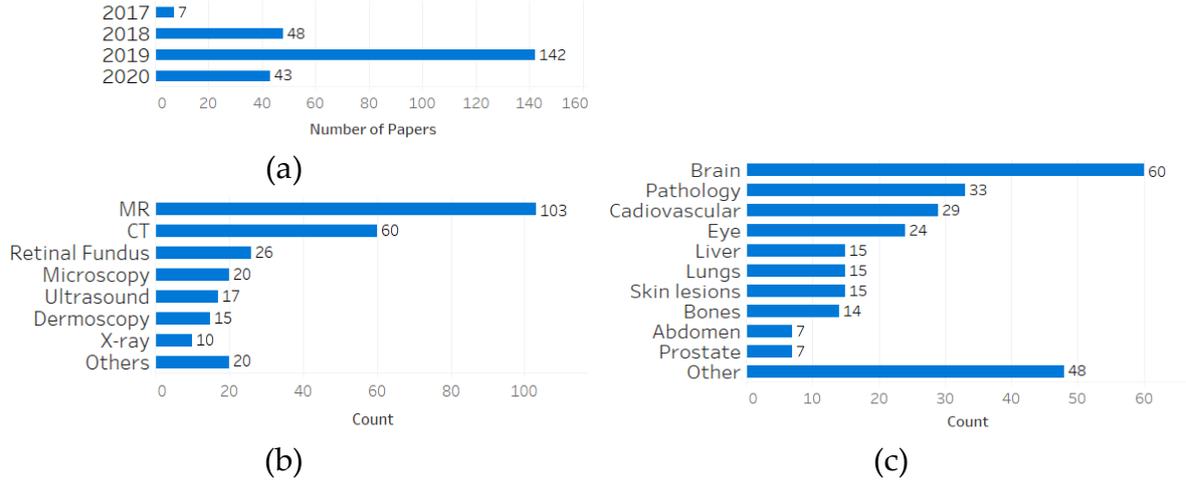

**Figure 1.** (a) Distribution of U-net related papers in our survey by year of publication starting with 2017. (b) Distribution of image modality in U-net related papers. (c) Distribution of application area in U-net related papers. It should be noted that the statistic for 2020 is only till 04/01/2020. We expect the total volume of U-net related papers to be published in 2020 to be proportionally higher than the first three months. It should also be noted that some papers had multiple image modalities and application areas, and each instance was counted separately.

What makes U-net particularly useful is its creation of highly detailed segmentation maps using very limited trading samples. The latter trait is of great importance in the medical imaging community as properly labeled images are often limited. This is achieved by utilizing random elastic deformation on the training data which enables the network to learn these variations without requiring new labeled data [1]. Another challenge is to separate touching objects of the same class, which is resolved by applying a weighted loss function that penalizes the model if it fails to separate the two objects. Finally, U-net is also much faster to train than most other segmentation models due to its context-based learning.

Since its inception in 2015, U-net has seen an explosion in usage in medical imaging. And naturally, there have been many advancements in U-net architecture by researchers implementing new methods or incorporating other imaging methods to U-net. In this survey, we go through papers that make use of U-net implantations relating to medical image analysis. To avoid redundancy, only papers since 2017 were reviewed. As there are numerous sources of scientific publicization, in order to find the relevant quality of research papers, we limited ourselves to three major publishers, IEEE, Springer, and Elsevier, and searched their databases with related keywords to find the top papers in each database and collected the appropriate papers. Since new papers are being published regularly, a designated endpoint of 04/01/2020 was selected which was the most recent paper collected in our initial search.

## 2. U-net architectures

### 2.1 Base U-net

As mentioned earlier, the U-net network can be divided into two parts: The first part is the contracting path which uses a typical CNN architecture. Each block in the contracting path consists of two successive 3x3 convolutions followed by a ReLU activation unit and a max-pooling layer. This arrangement is repeated several times. The novelty of u-net comes in the expansive path where at each stage the feature map is upsampled using 2x2 up-convolution. Then, the feature map from the corresponding layer in the contracting path is cropped and concatenated onto the upsampled feature map. This is followed by two successive 3x3 convolutions and ReLU activation. At the final stage, an additional 1x1 convolution is applied to reduce the feature map to the required number of channels and produce the segmented image. The cropping is necessary since pixel features in the edges have the least amount of contextual information and therefore need to be discarded. This results in a network resembling a u-shape and more importantly, propagates contextual information along the network which allows it to segment objects in an area using context from a larger overlapping area.

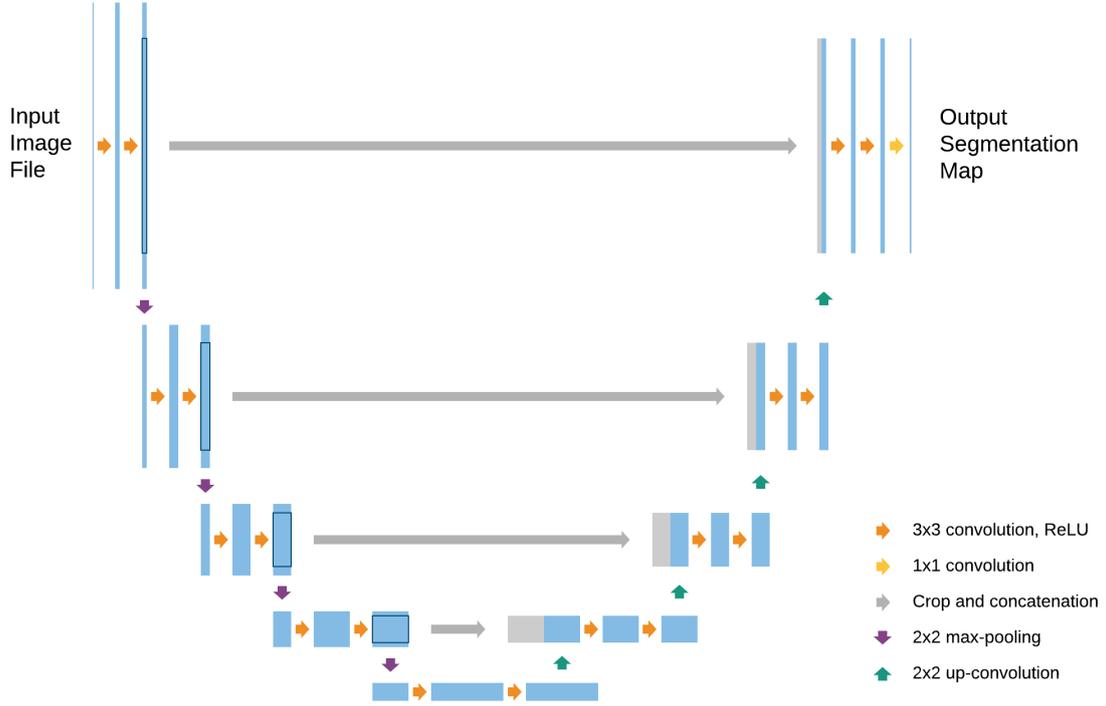

**Figure 2.** Basic U-net architecture. The arrows represent the different operations, the blue boxes represent the feature map at each layer, and the gray boxes represent the cropped feature maps from the contracting path.

The energy function for the network is given by the following equation:

$$E = \sum w(x) \log \left( p_{k(x)}(x) \right), \tag{1}$$

Where $p_k$ is the pixel-wise SoftMax function applied over the final feature map.

$$p_k(x) = \frac{e^{a_k(x)}}{\sum_{k'=1}^{K} e^{a_{k'}(x)}}, \tag{2}$$

And $a_k(x)$ denotes the activation in channel k.

### 2.2 3D U-net

3D U-net is an augmentation of the basic U-net framework that enables 3D volumetric segmentation [3]. The core structure is the same having both a contracting and expansive path. However, all of the 2D operations are replaced with corresponding 3D operations, namely 3D convolutions, 3D max pooling, and 3D up-convolutions resulting in a 3-dimensional segmented image. This network is able to segment images using very few annotated examples. This is due to the fact that 3D images have lots of repeating structures and shapes, enabling a faster training process even with scarcely labeled data. 3D U-net has seen extensive use in volumetric CT and MR image segmentation applications, including diagnosis of the cardiac structures [4]–[11], bone structures [12]–[15], vertebral column [16], [17], brain tumors [18]–[20], liver tumors [21]–[23], lung nodules [24], nasopharyngeal cancer [25], multi-organ segmentation [26]–[28], head and neck organ at risk assessment [29], and white matter tracts segmentation [30]. 3D U-net has been used to great effect in many biomedical applications. Zeng et al. [12] created a network that produced multi-level segmented images allowing greater abstraction when making a diagnosis.

## 2.3 Attention U-net

An often-desirable trait in an image processing network is the ability to focus on specific objects that are of importance while ignoring unnecessary areas. The attention U-net achieves this by making use of the attention gate [31], [32]. An attention gate is a unit which trims features that are not relevant to the ongoing task. Each layer in the expansive path has an attention gate through which the corresponding features from the contracting path must pass through before the features are concatenated with the upsampled features in the expansive path. Repeated uses of the attention gate after each layer improves segmentation performance significantly without adding too much computational complexity to the model.

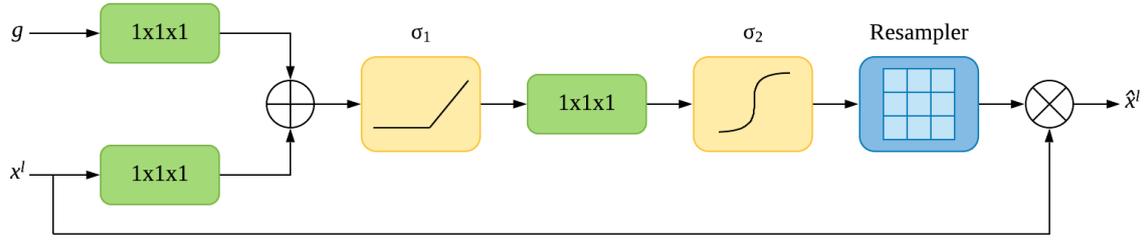

**Figure 3.** Additive attention gate schematic. The input signal $x^l$ and the gating signal $g$ both pass through separate 1x1x1 convolutions. The signals are then added and undergo a series of linear transformation which are ReLU activation ($\sigma_1$), a 1x1x1 convolution, sigmoid activation ($\sigma_2$), and an optional grid resampler. Finally, the original input is concatenated to the output from the sigmoid unit or the resampler.

The attention unit is useful in encoder-decoder models such as the U-net since it can provide localized classification information as opposed to global classification. In U-net this allows different parts of the network to focus on segmenting different objects. Furthermore, with properly labeled training data, the network can attune to particular objects in an image. The attention gate applies a function where the feature map is weighted according to each class and the network can be tuned to focus on a particular class [33], and hence pay attention to particular objects in an image. While there are different types of attention gates, additive attention is more popular in image processing due to it resulting in higher accuracy. The additive attention gate is described by the following:

$$q_{att}^l = \psi^T \left( \sigma_1 \left( W_x^T x_i^l + W_g^T g_i + b_g \right) \right) + b_\psi, \tag{3}$$

$$\alpha_i^l = \sigma_2(q_{att}^l(x_i^l, g_i; \Theta_{att})) \gamma \tag{4}$$

where $x^l$ is the features from the contracting path and $g$ is the gating signal. The term $\sigma_2(x_{i,c})$ represents the sigmoid function:

$$\sigma_2\left( x_{i,c} \right) = \frac{1}{1 + \exp\left( -x_{i,c} \right)}, \tag{5}$$

Attention U-net has been applied on problems such as ocular disease diagnosis [34]–[38], melanoma [39], lung cancer [34], cervical cancer [40], abdominal structure segmentation [32], fetus development [32], and brain tissue quantification [41].

## 2.4 Inception U-net

Most image processing algorithms tend to use fixed-size filters for convolutions. Tuning the model to find the correct filter size can often be cumbersome. Moreover, fixed-size filters are appropriate only for images with similar size salient parts. In many applications, the analysis looks through images with large variations in shapes and sizes in the salient region. One solution to this problem would be to use deeper networks that can read high-level details across a spectrum of sizes and shapes. However, such deep networks are quite computationally expensive. An alternative solution, called the inception network, uses filters of multiple sizes on the same layer in the network. [42]. The outputs from the different filters are concatenated and transferred onto the next layer. The inception network is able to analyze images with different salient regions quite effectively due to the different filter sizes. To reduce computational complexity, the inception network adds a 1x1 convolution before every 3x3 or larger filter for dimensionality reduction. Additionally, pooling layers may also be added parallel in each inception module.

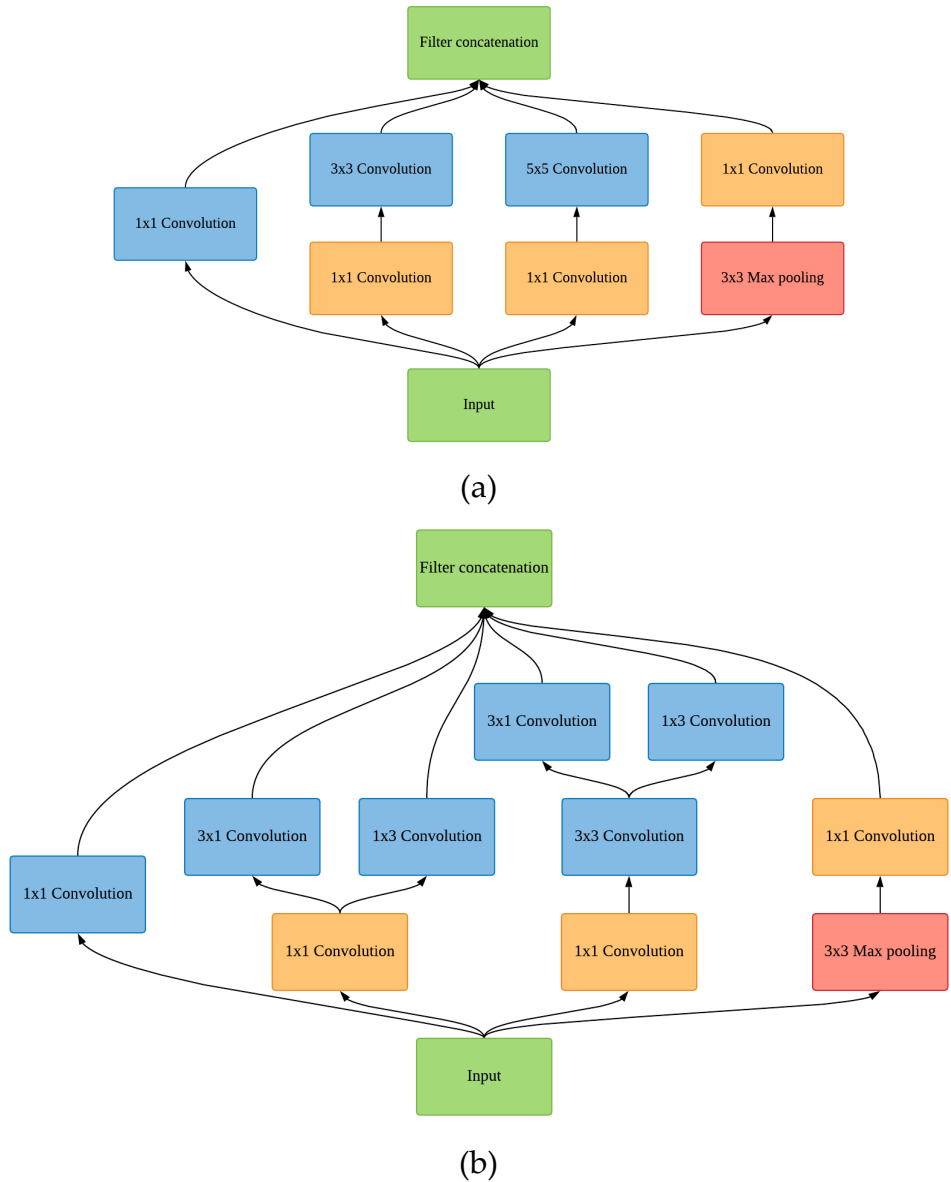

(a)

(b)

**Figure 4.** (a) The original inception block used in GoogLeNet. (b) Improved inception block with factorized filters. At the end of the inception block, the feature maps from each filter are concatenated together and passed onto the next layer. It should be noted that both networks in figures (a) and (b) are equivalent, though the factorized network requires less computational power.

The original inception network, called GoogLeNet, attained the state of the art outcomes in the ILSVRC14 competition [42]. Soon after, however, more improvements to the network were made. To further improve performance, factorization methods were applied. 5x5 convolution was replaced with two successive 3x3 convolutions. A single 5x5 convolution is 2.78 times more computationally expensive than two equivalent 3x3 convolutions [43]. Further factorization can be applied by splitting nxn filters into a 1xn and nx1 filter respectively. Factorizing a 3x3 filter by this method makes the network 33% less expensive.

Inception modules of different configurations have been applied on a multitude of U-net applications including brain tumor detection [19], [44], [45], brain tissue mapping [46], cardiac segmentation [7], [47], lung nodule detection [48], human embryo segmentation [49], and ultrasound nerve segmentation [50].

*2.5 Residual U-Net*

This variant of U-net is based on the ResNet [51] architecture. The motivation behind ResNet was to overcome the difficulty in training highly deep neural networks. It is known that neural networks are able to converge faster to a solution the more layers are present. However, experimental results showed that increasing the number of layers results in saturation, and further increases can cause degradation of performance [51]. This degradation arises due to the loss of feature identities in deeper neural networks caused by diminishing gradients in the weight vector. ResNet lessens this problem by utilizing skip connections which take the feature map from one layer and add it to another layer deeper in the network. This behavior allows the network to better preserve feature maps in deeper neural networks and provide improved performance for such deeper networks.

In the residual U-net, at each block in the network, the input to the first convolutional layer is added to the output from the second convolutional layer using a skip connection. This skip connection is applied before the down-sampling or upsampling in the corresponding paths in the U-net. The usage of residual skip connections helps alleviate the vanishing gradient problem [51] allowing for U-net models with deeper neural networks to be designed. Each residual unit can be denoted by the following expressions:

$$y_l = h(x_l) + \mathcal{F}(x_l, \mathcal{W}_l) \tag{6}$$

$$x_{l+1} = f(y_l) \tag{7}$$

Here $x_l$ and $x_{l+1}$ correspond to the input and output of the residual unit, $F(\cdot)$ corresponds to the residual function, $f(\cdot)$ is the activation function, and $h(\cdot)$ is the identity mapping function.

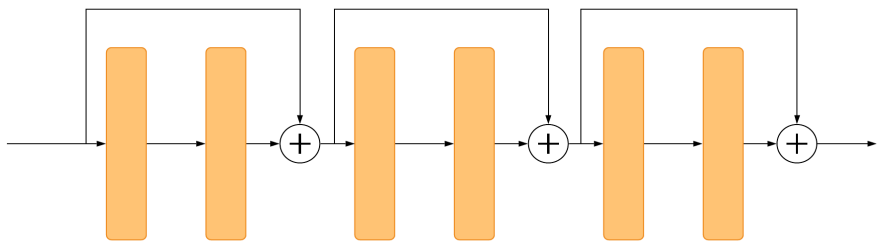

**Figure 5.** Three successive ResNet blocks with skip connections. The skipped signal is joined with the channel output via element-wise addition. The most common ResNet implementations are double layer skips (as shown in this figure) or triple layer skips.

We have found papers where deep residual U-nets have been used to great effect in many biomedical imaging applications such as nuclei segmentation [52], [53], brain tissue quantification [41], brain structure mapping [54], retinal vessel segmentation [55], breast cancer [56], liver cancer [23], [57], prostate cancer [58], endoscopy [59], melanoma [59], osteosarcoma [60], bone structure analysis [61], and cardiac structure analysis [58], [62]. Deep residual U-nets are ideal for complex image analysis.

*2.6 Recurrent Convolutional Network*

Recurrent neural networks are a type of neural network initially designed to analyze sequential data such as text or audio data. The network is designed in such a way so that a node's output is changed based on the previous output from the same nodes; i.e. a feedback loop as opposed to a traditional feedforward network. This feedback loop creates an internal state or memory that provides the node with temporal properties that change the output at discrete time steps. When extended to the entire layer, this allows the network to process contextual information from the preceding data.

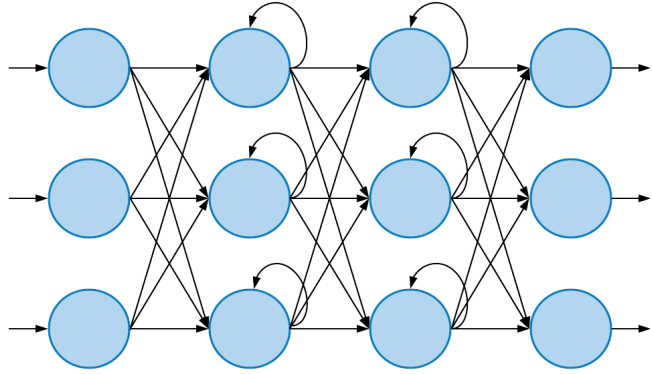

**Figure 6.** Recurrent neural network. In this simple network, the second and third layers are recurrent layers. Each neuron in a recurrent layer receives the feedback from its output as well as new information from the previous layer at discrete time periods and correspondingly produces a new output. This component allows the network to process sequential information.

The recurrent convolutional neural network (RCNN) [63] incorporates the recurrent feedback loops into a convolutional layer. The feedback is applied after both convolution and an activation function and feeds the feature map produced by a filter back into the associated layer. The feedback property allows the units to update their feature maps based on context from adjoining units, providing better accuracy and performance. The output *y* of the recurrent convolutional neural network can be expressed as below:

$$y_{ijk}^l(t) = \left(w_k^f\right)^T x_i^{f^{(i,j)}}(t) + (w_k^r)^T x_i^{r^{(i,j)}}(t-1) + b_k,$$ (8)

In this expression $x^f(t)$ is the feedforward input and $x^r(t-1)$ is the recurrent input for the l[th] layer, $w^f$ is the feedforward weight, $w^r$ is the recurrent weight, and *b* is the bias of the k[th] feature map. Recurrent convolutional layers have been used in [64], [65]. Alom et al. [52], [66] devised a U-net model containing both recurrent convolution layers and residual connections. The resulting network outperformed solely residual and recurrent U-net models as well as prior state of the art methods using a similar number of parameters.

*2.7 Dense U-net*

Dense U-nets employ DenseNet [67] blocks in place of regular layers. While the ResNet model allows for deeper neural networks, it does not eliminate the problem of vanishing gradients. ResNet architecture also eventually degrade in performance with increasing layers. DenseNet is a deep learning architecture built on top of ResNet with two key changes. Firstly, every layer in a block receives the feature or identity map from all of its preceding layers [67]. And the second major change is that the identity maps are combined via channel-wise concatenation into tensors [67] as opposed to ResNet where the identity maps are summed via element-wise addition. Therefore, the identity mapping of each layer is dependent not only on the previous layer but on all of the layers before it in the block. This allows DenseNet to preserve all identity maps from prior layers and significantly promote gradient propagation. The implication is that each layer can have fewer channels as information is more easily preserved between layers resulting in higher accuracy with fewer computations and as a consequence can allow deep learning models with a greater number of layers. The output for each layer in a dense block is described below:

$$x_l = H_l([x_0, x_1, x_2, \ldots, x_{l-1}]), \tag{9}$$

Where $H_l(\cdot)$ represents the dense mapping function which usually includes batch normalization, ReLU activation, and a convolutional layer while [] denotes channel-wise concatenation.

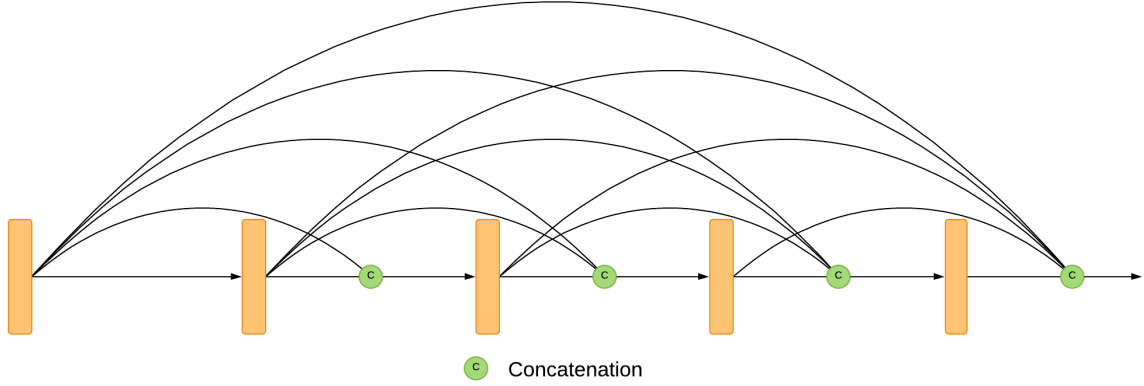

**Figure 7.** A five-layer dense block. The concatenation unit receives the feature map from all previous layers and passes it onto the next layer. This ensures that any given layer has contextual information from any of the previous layers in the block.

When implementing a U-net, each traditional U-net block is replaced with a dense block of two or more convolutional layers. The adoption of dense blocks allows for deeper U-net models which can segment objects in an image with greater distinction. This attribute of dense U-nets is highly sought after in medical image analysis due to objects in such images being extremely close and often overlapping. Applications of dense U-net have been found in analysis of brain tumors [20], [45], retinal blood vessel segmentation [45], cerebral blood vessel segmentation [68], [69], melanoma [70], lung cancer [70], liver cancer [71], and multi-organ segmentation [72].

### 2.8 U-net++

U-net++ is another powerful form of the U-net architecture inspired from DenseNet [67]. It uses a dense network of skip connections as an intermediary grid between the contracting and expansive paths [73]. This aids the network by propagating more semantic information between the two paths enabling it to segment images more accurately.

In traditional U-net, the feature maps of the contracting path are directly concatenated onto the corresponding layers in the expansive path. U-net++ however has a number of skip connection nodes between each corresponding layer. Each skip connection unit receives all of the feature maps from all previous units at the same level as well an upsampled feature map from its immediate lower unit. Therefore, each level is equivalent to a dense block. This arrangement minimizes the loss of semantic information between the two paths. The operation of the skip connection unit is as follows, where $x$ represents the feature map and $i$ and $j$ correspond to the index down the contracting path and across the skip connections respectively:

$$x^{i,j} = \begin{cases} \mathcal{H}(x^{i-1,j}), & j = 0 \\ \mathcal{H}([[x^{i,k}]_{k=0}^{j-1}, \mathcal{U}(x^{i+1,j-1})]), & j > 0 \end{cases}' \tag{10}$$

Here $H(\cdot)$ denotes a convolution and activation operation, $U(\cdot)$ represents the up-sampling operation, and [] signifies a concatenation. The number of intermediary skip connection units depends on the layer number and decreases linearly going down the contracting path. Applications in U-net++ include segmentation of cell nuclei [73], cancer tissue [73], cardiac structures and vessels [74], [75], and pelvic organs [76].

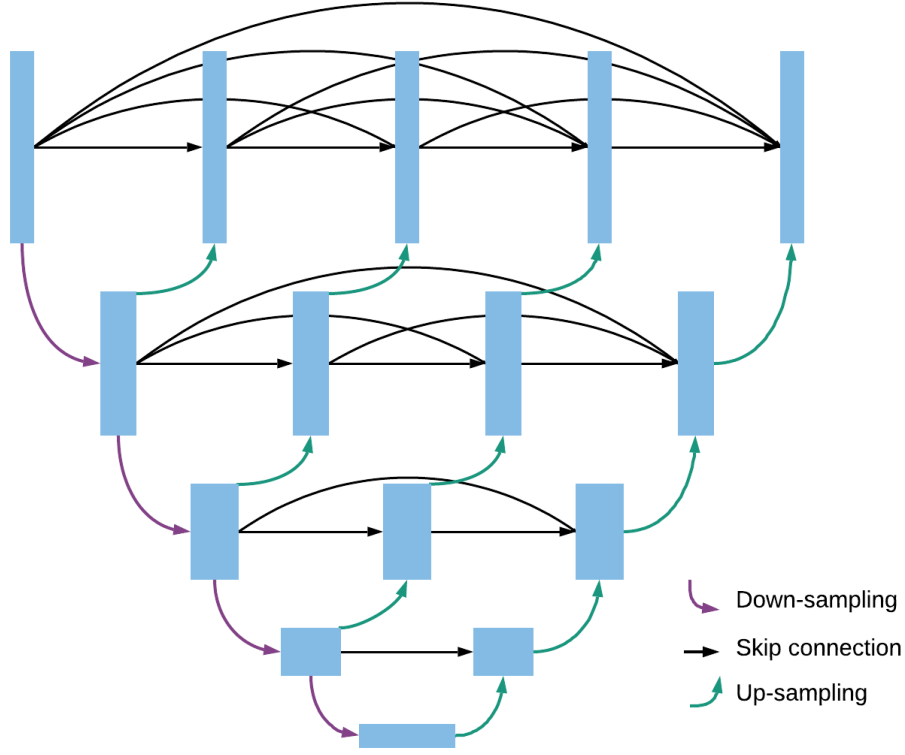

**Figure 8.** U-net++ schematic representation. Each square denotes a convolutional block. Unlike base U-net, which has a single direct concatenation from the contracting path to the expansive path, U-net++ has a series of intermediary convolutional blocks between the two paths. Each intermediary and expansive block receives the concatenated feature maps from all of the previous blocks at the same level as well as the upsampled feature map from the block immediately below it.

*2.9 Adversarial U-net*

An adversarial model is a setup where two networks compete against each other in order to improve their performance. Generative adversarial networks (GAN) are a novel type of adversarial process used to generate new data [77]. The framework consists of two networks: a discriminator and a generator. The discriminator network *D*, is a classifier that is trained to identify whether a given input image came from the data set or produced by the generator *G*. The discriminator *D* undergoes standard CNN supervised training and for each image input, it outputs the probability of the image being produced by *G* with the goal of minimizing its error rate of classifying 'fakes' as real data set images. The generator *G* produces images that are periodically fed to the discriminator. To help the generator produce convincing images, its gradient function is a function of the discriminator's gradient function during the step the discriminator is fed a fake image. This allows the generator to adjust its weight in response to the output of the discriminator. Furthermore, to create variations in the images produced by the generator, random noise is passed to it. The goal of the generator is to deceive the discriminator, i.e. maximize the error rate of the discriminator. This minimax relationship results in an adversarial network where the two networks compete with each other.

$$\min_G \max_D V(D, G) = \mathbb{E}_{x \sim p_{data}(x)} \log D(x) + \mathbb{E}_{z \sim p_z(z)} \log \left(1 - D\big(G(z)\big)\right) \tag{11}$$

Given enough time the adversarial network should reach an optimal state where the discriminator always outputs a probability of ½ regardless of whether the image is from the data set or the generator [77], meaning it can no longer distinguish the real images from the synthetic images produced by the generator. The resulting generator can then be used to artificially create images of a particular subject.

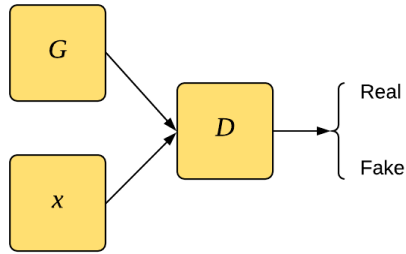

**Figure 9.** GAN block diagram. The goal of network *D* is to classify all inputs from *x* and network *G* as real and fake respectively. The goal of *G* is to have its output evaluated as real.

This framework can be further extended to restrict the GAN into producing a limited band of synthetic images by controlling its labels and input images. This alteration is known as a conditional GAN [78]. Adversarial U-nets are a type of conditional GANs. The generator network is constructed based on the U-net architecture while the discriminator remains the same. The U-net design allows the generator to take an image as input instead of random noise. The key difference in adversarial U-nets is that the goal of the generator is not to produce new images, but rather transformed images. This output of the generator *G* is evaluated against the discriminator *D* which is trained on manually transformed images. Ideally, after proper training, the generator will be able to achieve the same transformation ability as the manual human transformation. The resulting generator can then be used to apply its transformation function on new images which would be considerably faster than a physician manually converting the image. Adversarial U-nets have seen a wide spectrum of applications including quantitative susceptibility mapping of the brain [79], detection of brain tumors [80], [81] and breast cancer [82], segmentation of retinal vessels [38], segmentation of cardiac structures [83], and image registration of brain structures [84].

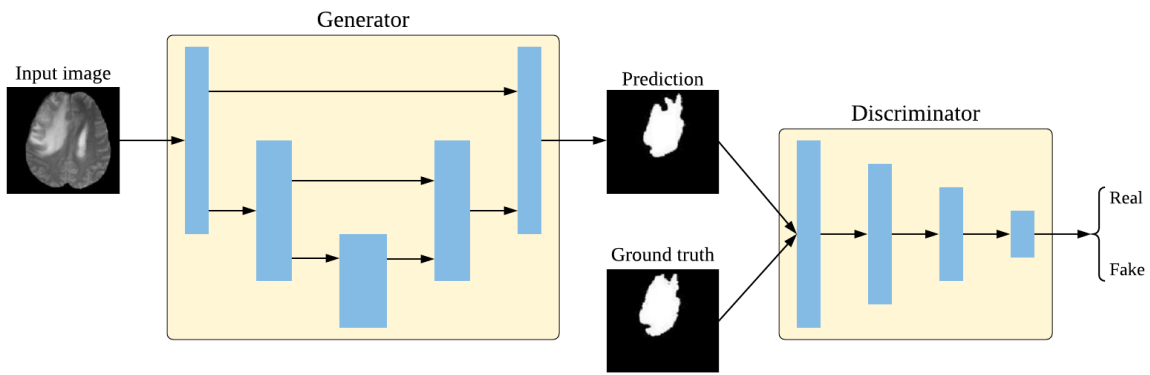

**Figure 10.** Simplified schematic of U-net based GAN. The generator synthesizes predictions for the tumor area from the input images. The predictions are fed into the discriminator which judges the accuracy of the prediction by evaluating its similarity to the ground truth. If the prediction is similar to the ground truth then the discriminator will be unable to distinguish between them and classify the prediction as real. Given enough training, the GAN will be able to segment images to the same accuracy and precision as manual annotations [81].

*2.10 Cascaded arrangement*

In addition to the aforementioned architectures, various other network configurations have also been tested for U-net. One such method is cascading two or more U-nets. In this arrangement, the first layer performs a high-level segmentation, with each successive U-net performing segmentation on smaller objects. Feng et al. [85] designed a two-stage U-net model where the first U-net segments the liver from other organs and the second U-net segments tumors within the liver. Liu et al. [57] designed a two-stage U-net for liver segmentation with an intermediate processing module between the two U-nets. Xu et al. [8] and Li et al. [44] have both designed two-stage cascaded U-nets where the first network is a 2D U-net and the second network is a 3D U-net. Other two-stage U-net models are implemented in [5], [6], [29], [53], [71], [86]–[89]. While two-stage networks are the most common type of cascaded U-nets, we have found two instances of cascaded U-nets with variable numbers of stages [90], [91]. In all of these papers, the cascaded U-net performed better than a single U-net.

*2.11 Parallel arrangement*

Yet another arrangement of the overall architecture can be found in the form of a parallel arrangement of part or the entirety of a U-net network. Abd-Ellah et al. [92] arranged two parallel U-nets and aggregated the results for improved segmentation accuracy. Soltanpour et al. [93] implemented four parallel U-nets with each segmenting a different CT map and then merging the results. A halfway point can be achieved by parallel encoders, which allow for better extraction of features [94]–[96]. Murugesan et al. [97] implemented a network with parallel decoders that provide different levels of segmentation.

2.5D U-net is a special architecture where three 2D U-net networks are run parallelly on different 2D projections of a 3D image to produce a 3D segmentation map. The 2D U-nets perform slice by slice segmentation on the 3D volume along three different axes, and the final 3D segmentation is computed by fusing the results [98]–[101]. The advantage of the 2.5D parallel arrangement is reduced computational load for segmentation when juxtaposed with an equivalent 3D network.

## 3. Image modalities

Segmentation is the primary task for U-net models. The goal of segmentation tasks is to outline and separate different objects in an image, i.e. to classify different objects rather than classifying the whole image. This is of particular importance in the medical imaging community as the diagnosis of medical conditions requires careful analysis of local regions in an image. For instance, the diagnosis of brain tumors would require separating the tumors from the rest of the brain structures. We have found extensive use of the U-net architecture for an assortment of medical imaging analysis. In the next section, we discuss the major image modalities on which U-net has been applied.

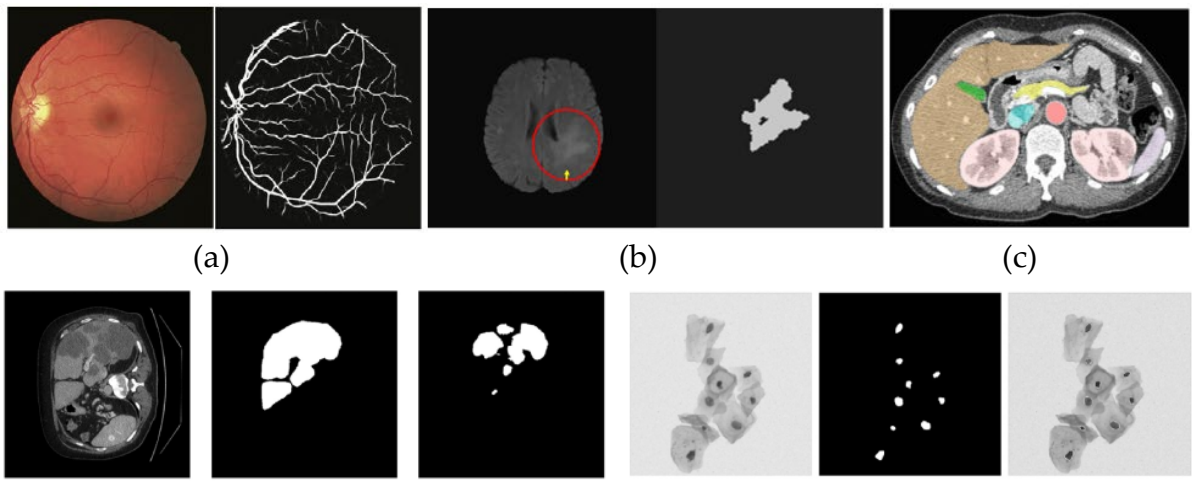

(a)                                      (b)                                      (c)

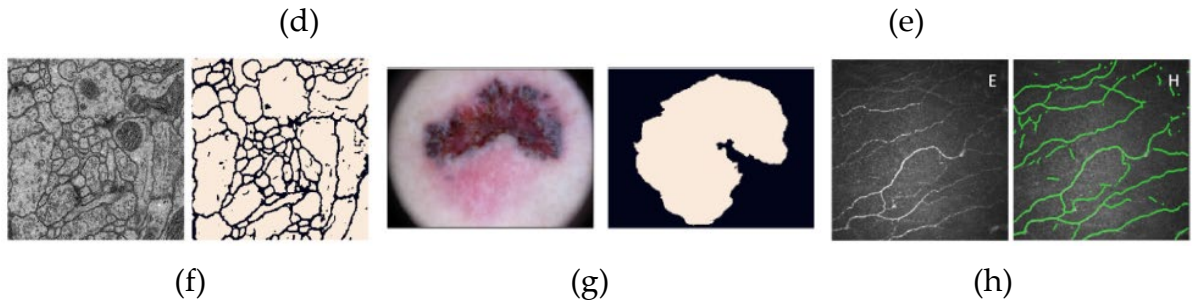

**Figure 11.** Examples of U-net applications. Images have been collected from papers in this survey. (a) Retinal vessel segmentation [102]. (b) Brain tumor detection and segmentation [103]. (c) Multi-organ abdominal segmentation (liver; spleen; left and right kidneys; pancreas; gallbladder; aorta; and inferior vena cava) on CT scans [27]. (d) Liver tumor segmentation, left to right: original CT image, liver segmentation image, and lesion segmentation image [104]. (e) Nuclei prediction, from left to right: original cell images, prediction of nuclei, labeling nuclei in the original images [40]. (f) Cell segmentation [59]. (g) Skin lesion segmentation [59]. (h) Corneal nerve segmentation [105].

### 3.1 Magnetic resonance imaging (MRI)

MRI is a very popular radiology imaging technique used to take pictures of soft tissue inside the body. In our review, we have found MRI to be the most popular image modality for segmentation using U-net. MRI is a useful diagnostic tool particularly for analysis of the brain. U-net has been used extensively in this regard for the segmentation of brain structures. Various U-net models have been applied on MR images for brain tumor diagnosis [18], [19], [59], [45], [81], [106], [107], [87], [108]–[111], [92], [112]–[119], [44], [120]. U-net has also been applied on brain tissue for investigation of neurological conditions [54], [101], [103], [121]–[124], analysis of white matter tissue [125]–[127], [89], fetal brain development [128]–[131], and stroke lesions [69], [132]–[136].

U-net implementation has also been applied on cardiovascular MR images [4]–[6], [8]–[11], [47], [58], [62], [74], [83], [86], [90], [137]–[146] to segment structures of the heart. Cancer is a leading cause of deaths worldwide and MR is one of the strongest methods for proper prognosis of various types of cancers. In addition to brain cancer, we have found applications on prostate cancer [58], [64], [147]–[151], liver cancer [21], [152], [153], nasopharyngeal cancer [25], [98], [154], and breast cancer [99], [155]. Other implementations include segmentation of the femur [12]–[14], spinal cord [156], [157], blood vessels [100], vertebral column [17], human placenta [158], and the uterus [159].

**Table 1.** Applications of U-net based models for MR image analysis.

| Reference | Model/Methods used |
|---|---|
| **Brain tumor** | |
| [106], [107], [110]–[112], [114]–[118] | Base U-net |
| [18], [109], [120] | 3D U-net |
| [81] | Adversarial net; GAN |
| [59], [108] | Residual block |
| [113] | Dense block |
| [87] | Cascaded U-net |
| [92] | Residual block; Parallel U-net |
| [44] | Inception block; Up skip connections |
| [45] | Dense block; Inception block |
| [119] | 3D U-net; Residual block |
| [19] | 3D U-net, Inception block, Residual block |
| **Brain tissue** | |
| [103], [121]–[124] | Base U-net |
| [28], [160] | 3D U-net |
| [161] | 2.5D U-net |
| [54] | Residual block |
| [101] | Parallel U-net |
| [41] | Attention gate; Residual block |
| **White matter tracts** | |
| [126], [127] | U-net with modified skip connections |
| [125] | Base U-net |
| [89] | Cascaded U-net |
| **Fetal brain** | |
| [128]–[130] | Base U-net |
| [131] | Base U-net; 3D U-net |
| **Stroke lesion/thrombus** | |
| [133]–[136] | Base U-net |
| [132] | 3D U-net |
| [69] | Dense block; Inception block |
| **Cardiovascular structures** | |
| [138], [140]–[142], [144], [146] | Base U-net |
| [62], [139], [143], [145] | Residual block |
| [4], [9], [10], [28] | 3D U-net |
| [86], [90] | Cascaded U-net |
| [5], [8] | Cascaded 3D U-net |
| [11] | Base U-net; 3D U-net |
| [83] | Adversarial net; GAN |
| [58] | Residual block |
| [137] | Dense block |
| [74] | U-net++ |
| [47] | Inception block; Residual block |
| [6] | Cascaded 3D U-net; Residual block |
| **Prostate cancer** | |
| [147], [149]–[151] | Base U-net |
| [28] | 3D U-net |
| [64] | Recurrent net |
| [58] | Residual block |
| **Liver cancer** | |
| [152], [153] | Base U-net |
| [21] | 3D U-net |
| **Nasopharyngeal cancer** | |
| [25] | 3D U-net; Residual block |
| [98] | Parallel U-net |
| [154] | Modified convolution block |
| **Femur** | |
| [12]–[14] | 3D U-net |
| **Breast cancer** | |
| [155] | Base U-net |
| [99] | Parallel U-net |
| **Spinal cord** | |
| [156], [157] | Base U-net |
| **Blood vessels** | |
| [100] | Base U-net |
| **Placenta** | |
| [158] | Base U-net |
| **Uterus** | |
| [159] | Base U-net |
| **Vertebral column** | |
| [17] | 3D U-net |

*3.2 Computed tomography (CT)*

CT scans are another major non-invasive medical analysis tool for analyzing internal organs and tissue. As with MRI, cancer diagnosis is a major area where CT imaging is applied; liver cancer [22], [23], [57], [71], [73], [85], [94], [104], [162]–[164], lung cancer [24], [34], [45], [48], [70], [73], [165]–[169], bone cancer [60], and cervical cancer [170]. CT scans are also used for multiorgan abdominal segmentation [26], [27], [32], [72], [151], [171]. CT scans have also been used for the segmentation of hard tissue such as bones [14]–[16], [60], [76], [172]. Along with MR imaging, CT is one of the few imaging techniques that can produce 3D images. The versatility of CT imaging makes a favored modality in medical diagnosis.

**Table 2.** Applications of U-net based models for CT image analysis.

| Reference | Model/Methods used |
| --- | --- |
| Liver cancer | |
| [104], [162], [163] | Base U-net |
| [22], [28] | 3D U-net |
| [23] | 3D U-net; Residual block |
| [73] | U-net++ |
| [85] | Cascaded U-net |
| [57] | Cascaded U-net; Residual block |
| [71] | Cascaded U-net; Dense block |
| [94] | Modified U-net with dual parallel encoders |
| Lung cancer | |
| [165]–[167] | Base U-net |
| [168], [169] | Residual block |
| [34] | Attention gate |
| [24] | 3D U-net; Residual block |
| [48] | Dense block; Inception block |
| [73] | U-net++ |
| Pulmonary tissue | |
| [173], [174] | Base U-net |
| [175] | Residual block |
| Abdominal organs | |
| [151], [171] | Base U-net |
| [26], [27] | 3D U-net |
| [32] | Attention gate |
| [72] | Dense block |
| Cardiovascular structures | |
| [4] | 3D U-net |
| [86] | Cascaded U-net |
| [5] | Cascaded 3D U-net |
| [119] | 3D U-net; Residual block |
| [7] | 3D U-net; Inception block |
| [75] | U-net++ |
| Pancreas | |
| [156], [176], [177] | Base U-net |
| [28] | 3D U-net |
| Bones | |
| [172] | Base U-net |
| [14] | 3D U-net |
| [15] | 3D U-net; Residual block |
| [60] | Residual block |
| [76] | U-net++ |
| Stroke lesions | |

| Reference | Model/Methods used |
|---|---|
| [93], [178] | Base U-net |
| [132] | Base U-net; 3D U-net |
| [69] | Dense block; Inception block |
| Head and neck | |
| [179] | 3D U-net |
| [29] | Cascaded 3D U-net |
| Gallstones | |
| [180] | U-net++ |
| [88] | Cascaded U-net |
| Liver and spleen | |
| [164] | Dense block |
| Blood vessels | |
| [181] | Residual block |
| Brain | |
| [46] | Inception block; Residual block |
| [95] | Modified U-net with dual parallel encoders |
| Cervical cancer | |
| [170] | Base U-net |
| Fetus | |
| [32] | Attention gate |
| Melanoma | |
| [70] | Dense block |
| Vertebral column | |
| [16] | 3D U-net |

## 3.3 Retinal fundus imaging

Color fundus imaging is an ophthalmology technique used for the detection and diagnosis of various ocular diseases such as glaucoma, diabetic retinopathy, age-related macular degeneration (AMD), etc. Proper prognosis depends on the precise segmentation of key structures such as retinal blood vessel segmentation [55], [182]. Accurate screening is of chief importance since such diseases often need to be diagnosed early for treatment. Though ophthalmic imaging has a far narrower scope than MR and CT, the retinal fundus is one of the most analyzed structures in our survey right after the brain and cardiovascular system. As it is the primary method of imaging the retina, we expect more research on fundus image analysis to continue as well as research on more complex retinal fundus images.

**Table 3.** Applications of U-net based models for fundus image analysis.

| Reference | Model/Methods used |
|---|---|
| [102], [182]–[190] | Base U-net |
| [34]–[36], [191] | Attention gate |
| [55], [169], [192], [193] | Residual block |
| [70] | Dense block |
| [38] | Adversarial net; GAN; Attention gate |
| [91] | Cascaded U-net |
| [194] | Attention gate; Residual block |
| [45] | Dense block; Inception block |
| [195] | Inception block; Residual block |
| [97] | Modified U-net with parallel decoders |
| [196] | Recurrent residual block; Up skip connections |

## 3.4 Microscopy

Microscopy refers to the examination of extremely small objects that cannot be observed with the naked eye. It should be noted that in our survey we refer to microscopy to mean only optical microscopy. This modality is used extensively in pathology. One of the major challenges in microscopy imaging is identifying overlapping cells as well as identifying the boundary between cells. These are unique challenges to microscopy as smaller structures such as cells and tissues often don't have well-defined landmarks and similarities that make it harder for image processing. However, U-net has overcome such challenges [1] and continues to be a strong implementation for this modality.

**Table 4.** Applications of U-net based models for microscopy image analysis.

| Reference | Model/Methods used |
|---|---|
| Cell nuclei | |
| [197] | Base U-net |
| [59], [198], [199] | Residual U-net |
| [52] | Recurrent net; Residual block |
| [53] | Cascaded U-net; Residual block |
| [73] | U-net++ |
| Cell contour | |
| [200], [201] | Base U-net |
| [40] | Attention U-net |
| Human embryo | |
| [202] | Base U-net |
| [49] | Inception block |
| Corneal nerve | |
| [105] | Base U-net |
| [36] | Attention gate |
| Chromosomes | |
| [203], [204] | Base U-net |
| Blood vessels | |
| [205] | Base U-net |
| Parasite detection in Chagas disease | |
| [206] | Base U-net |
| Sclerosis | |
| [207] | Base U-net |
| Colon gland | |
| [208] | Base U-net |

*3.5 Dermoscopy*

Dermoscopy is a detailed examination of the skin. It is almost exclusively used to examine skin diseases such as skin lesions. The primary medical condition diagnosed using Dermoscopy images in our survey is melanoma or skin cancer, though we have found a single paper on psoriasis diagnosis [209]. The performance of Dermoscopy image analysis methods is of keen interest in the medical imaging community since it is often used for early detection of melanoma and is less costly than other noninvasive diagnostic tools.

**Table 5.** Applications of U-net based models for Dermoscopy image analysis.

| Reference | Model/Methods used |
|---|---|
| Melanoma | |
| [210]–[218] | Base U-net |
| [39] | Attention gate |
| [59] | Residual block |
| [70] | Dense block |
| [194] | Attention gate; Residual block |
| [196] | Up skip connections |
| Psoriasis | |

| [209] | Base U-net |

## 3.6 Ultrasound

Medical ultrasound is yet another noninvasive imaging technique for the analysis of internal structures. Ultrasound is mostly used for early and real-time diagnosis. Additionally, unlike many other image modalities, ultrasound devices are more maneuverable and can capture images from multiple angles. Ultrasound is also safe since it does not use radiation, hence it is the primary imaging modality for pregnancy-related diagnosis [219]–[221]. Medical ultrasound use cases also include analysis of soft tissue such as nerve bundles [39], [50], [151], [222], [223]. Its real-time image capture makes it extremely useful in tracking objects [96].

**Table 6.** Applications of U-net based models for ultrasound image analysis.

| Reference | Model/Methods used |
| --- | --- |
| Nerve segmentation | |
| [151], [222] | Base U-net |
| [50] | Inception block |
| [223] | Residual block |
| [224] | Modified parallel U-net |
| Breast lesion | |
| [225] | Base U-net |
| [39] | Attention gate |
| Arterial wall | |
| [226] | Base U-net |
| Cardiovascular structures | |
| [227] | Base U-net |
| Fetal head | |
| [219] | Base U-net |
| Gastrointestinal tumor | |
| [228] | Base U-net |
| Knee cartilage | |
| [96] | Modified U-net with dual parallel encoders |
| Preterm birth prediction | |
| [220] | Base U-net |
| Thyroid | |
| [229] | Residual block |
| Transcranial detection | |
| [221] | Base U-net |
| Ovary detection | |
| [230] | Base U-net |

## 3.7 X-ray

X-ray is a radiograph method used mainly for the imaging of hard tissue. It is the most widely used technique for the analysis of bones. U-net models have been applied to X-rays of bones for diagnosis of rheumatoid arthritis and osteoporosis [61], [231], and other bone-related diseases. Chest x-rays are also fairly prevalent and used for the detection of a myriad of pulmonary diseases including tuberculosis [194]. Aside from that, we have found applications of U-net in the detection of coronary stenosis [232], breast tumors [82], and surgical catheters [233].

**Table 7.** Applications of U-net based models for X-ray image analysis.

| Reference | Model/Methods used |
| --- | --- |
| Phalange bones | |
| [231], [234] | Base U-net |
| [61] | Residual block |
| Chest organs | |
| [235] | Base U-net |
| [34] | Attention gate |
| [194] | Attention gate; residual block |
| Pelvic bones | |
| [236] | Base U-net |
| Blood vessel segmentation | |
| [232] | Base U-net |
| Breast tumor | |
| [82] | Adversarial net; GAN |
| Surgical catheter detection | |
| [233] | Residual block |

### 3.8 Other Modalities

In addition to commonly used image modalities, we have also found u-net applications on more inconspicuous modalities. Endoscopy is an invasive imaging procedure where the imaging device is inserted into an organ or cavity to take pictures. U-net has been applied to endoscopy images for segmentation of polyps in the gastrointestinal tract [97], [237], colon objects [59], and detection of laryngeal leukoplakia [65]. On electron microscopy images, applications were detection of neuronal structures [156], [238], cell contour [156], and viruses [239]. Optical coherence tomography (OCT) is an imaging method for taking cross-sectional images of the retina. OCT is used for the diagnosis of various ocular diseases, for instance, age-related macular degeneration (AMD), retinal vein occlusion, and diabetic macular edema [240]. U-net has been used on OCT for segmentation of retinal layers [241], blood vessels [242], fluid regions [243], and Drusen [244]. Other uncommon applications are segmentation of blood vessels in digital subtraction angiography (DSA) [68], [245], white matter tract segmentation in diffusion tensor imaging [30], iris segmentation in iris imaging [37], and tumor detection in mammograms [56].

**Table 8.** Applications of U-net based models for various image modalities.

| Reference | Model/Methods used | Image Modality |
|---|---|---|
| [103], [237] | Base U-net | Endoscopy |
| [59] | Residual block | Endoscopy |
| [65] | Cascaded U-net; Recurrent residual net | Endoscopy |
| [97] | Modified U-net with parallel decoders | Endoscopy |
| [156], [238], [239] | Base U-net | Electron microscopy |
| [169] | Residual block | Electron microscopy |
| [240]–[244] | Base U-net | OCT |
| [245] | Base U-net | DSA |
| [68] | Dense block | DSA |
| [30] | 3D U-net | DTI |
| [37] | Attention gate | Iris imaging |
| [56] | Residual block | Mammogram |

## 4. Other Canonical Tasks by U-net

Even though U-net is an algorithm developed for segmentation, it has seen a modest amount of augmentation for other types of tasks. Image analysis is often hampered by the presence of noise or loss of detail during imaging. We have found three papers where U-net was implemented to remove artifacts from images by reconstructing the images [79], [246], [247] as well as a paper where U-net was used for de-aliasing [80]. Image registration is also an area where U-net model has seen experimentation [84], [248], [249]. Other reconstruction tasks include the correction of infant cortical surface [250] and EPID dosimetry correction of the cerebrospinal region [251]. Other outlier usages include synthesis of medical images [252], image super-resolution [20], and data augmentation to enable easier annotation of medical images [253].

**Table 9.** Applications of U-net based models on other canonical tasks.

| Reference | Image modality | Canonical task | Model/Methods | Application area |
|---|---|---|---|---|
| [246] | CT | Denoising | Modified U-net | Cervix |
| [247] | Ultrasound | Denoising | Base U-net | Brain tissue |
| [79] | MR | Denoising | 3D adversarial net | Brain tumor |
| [80] | MR | De-aliasing | Adversarial net | Brain tumor |
| [248], [249] | MR | Image registration | Base U-net | Brain tissue |
| [84] | MR | Image registration | Adversarial net | Brain tissue |
| [250] | MR | Image correction | 3D U-net | Brain surface |
| [251] | EPID dosimetry | Image correction | Base U-net | Brain and spinal cord |
| [252] | CT; MR | Image synthesis | Base U-net | Brain tissue |
| [253] | MR | Data augmentation | Base U-net | Brain tissue |
| [20] | MR | Superresolution | 3D U-net; Dense block | Brain tumor |

## 5. Discussion

Deep learning techniques such as U-net have seen increasing uses in medical image analysis over the years. Deep learning in image processing has allowed a variety of different tasks such as classification, detection, localization, etc. Segmentation tasks however are of keen interest in the medical imaging community. Surveys carried out by [254], [255] reveal that segmentation is the most sought out canonical task in medical image analysis. This is further evident by the abundance of papers published specifically for segmentation tasks, where U-net and its variants continue to be a prominent method [256]. The examination of U-net in this survey provides some answers to its high utility. In addition to being a well-performing segmentation model, a feature of U-net that makes it incredibly valuable is its high modularity and mutability. We have provided in this review numerous papers that have incorporated other deep learning methods into U-net, including some papers adopting multiple models at once. These alterations change the low-level architecture of the U-net while keeping the high-level design the same. More importantly, this means two things: the first is that this provides U-net a wide spectrum of applications since it can be greatly tuned depending on the application at hand. The second is that this means U-net still has a lot of potential for advancement since its modular nature allows it to keep on improving by incorporating newer novel ideas into itself.

For the specific use cases of the survey, we have found MR to be the most popular image modality though there remains a healthy variety of other image types. The same holds true for application areas, where U-net has been successfully implemented for both popular and niche applications. We would also like to point out to the alternate tasks performed in papers in this survey; although few, these tasks provide U-net with another avenue for exploration.

### 5.1 Challenges

The success of deep learning is vital for improved medical diagnosis. Although there has been tremendous progress in deep learning techniques such as U-net in the past decade, the nature of medical analysis demands algorithms to perform with minimal error. A major limitation of reducing this error in deep learning techniques is computational power. Powerful deep learning algorithms require more time to train and hence are less feasible. U-net algorithms have applied transfer learning as one solution to alleviate this problem [235]. EfficientNet is a framework for optimizing neural network construction that has the potential to streamline U-net design, making it more powerful using a similar number of parameters [257]. Another critical problem is the scarcity of annotated data for training. Ronneberger et al. [1] proposed a solution is his original U-net paper of applying random deformation to create new samples. An alternative solution is the use of adversarial models like GAN to synthesize new image samples. GAN in particular has seen tremendous success in synthesizing medical images [258]. Finally, deep learning models have the problem of being 'black boxes'; the input and output to the network are well understood, but the behavior of the internal hidden layers are not. This creates a problem where researchers often do not understand how to fix errors in the network or which layers or filters are more important to the task. Additionally, black boxes are difficult to interpret properly and their properties are difficult to replicate [259]. These are some key reasons why deep learning is yet to be used in any large scale real-world medical trial [260], despite its tremendous promise. However, day by day these problems are becoming easier to overcome and we expect to see even greater adoption of deep learning in the medical imaging community. In this regard, we expect U-net to be a major stepping stone in deep learning within the realm of medical image analysis.

### 5.2 U-net for COVID-19

The novel coronavirus (COVID-19) pandemic has created a staggering global medical crisis. As of June 22nd, a total of 8,708,008 cases and 461,715 deaths have been recorded globally [261]. To combat this challenge, the medical imaging community has involved itself in the research of various deep learning techniques, including U-net, for diagnosis of COVID-19. The primary diagnostic images taken for COVID-19 are chest CT scans which are ideal since U-net has seen extensive exploration in that modality. The versatility of the U-net network has allowed rapid development and deployment of early screening diagnostic algorithms for field use as early as March 2020 [262]. Further improvements on early screening tests have been made by augmenting attention and residual methods with U-net [263], [264]. Wu et al. [265] have implemented a hybrid network with U-net for segmentation and a classifier for classification. Yan et al. [266] developed a network with feature variation that allowed for an easier distinction of COVID-19 infection. U-net research has also been ongoing in X-ray based screening of COVID-19 [267], [268]. Alom et al. [269] established a multi-stage model to detect COVID-19 from X-ray and CT images. A survey on deep learning techniques for COVID-19 diagnosis reveals that U-net is one of the primary models of choice for segmentation related tasks [270]. This is no surprise as we have already explored the various utilities of U-net based models. We expect research on U-net based algorithms for the diagnosis of COVID-19 to continue and be a major asset to the medical imaging community during this global crisis.

## 6. Conclusions

In this survey, we hope to provide a starting point for researchers who wish to explore U-net, a powerful deep learning model used extensively for medical image segmentation. We explore the many variants of U-net and its diverse applications on a multitude of image modalities. We also breakdown the major deep learning methods and their application areas for all of the papers in this survey. It is concluded that U-net based architecture is indeed quite ground-breaking and valuable in medical image analysis. The growth of U-net papers since 2017 lends credence to its status as a premier deep learning technique in medical image diagnosis. Despite the many challenges remaining in deep learning-based image analysis, we expect U-net to be one of the major paths forward.

## 7. Resources

### 7.1 Frameworks

There are many open-source deep learning frameworks, among which some of the more popular and widely used frameworks are listed below:
- TensorFlow (Python, C, Java, Go, JavaScript, Swift): https://www.tensorflow.org/
- Keras (Python): https://keras.io/
- PyTorch (Python, C++): https://pytorch.org/
- Caffe (Python, MATLAB): http://caffe.berkeleyvision.org/
- Chainer (Python): https://chainer.org/
- Deeplearning4j (Java, Scala, Python, Clojure, Kotlin): https://deeplearning4j.org/

- Microsoft Cognitive Toolkit (CNTK) (Python, C#, C++):
  https://docs.microsoft.com/en-us/cognitive-toolkit/
- Theano (Python): http://www.deeplearning.net/software/theano/
- MXNet (Python, Scala, Julia, R, Clojure, Java, C++, Perl): https://mxnet.apache.org/
- ONNX (Python): https://microsoft.github.io/onnxruntime/
- Sonnet (Python): https://github.com/deepmind/sonnet
- PaddlePaddle (Python): https://github.com/PaddlePaddle/Paddle
- DeepGraphLibrary (Python): https://www.dgl.ai/

*7.2 SDKs*

- NVIDIA CUDA-X AI platforms: https://developer.nvidia.com/deep-learning-software
- Qualcomm mobile platforms: https://developer.qualcomm.com/solutions/artificial-intelligence

*7.3 Datasets*

The following includes some popular benchmarking datasets and databases for medical image segmentation tasks:
- ISBI 2012 cell segmentation challenge: Electron microscopy cell slices.
  http://brainiac2.mit.edu/isbi_challenge/

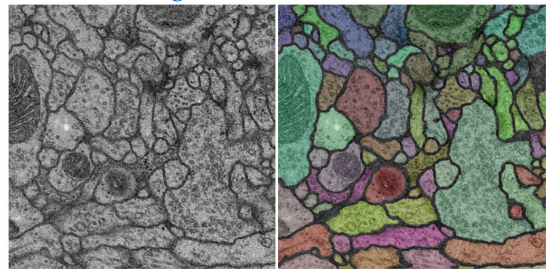

**Figure 12.** Example of ssTEM image and its corresponding segmentation from ISBI 2012 cell segmentation challenge [271].

- ISBI cell tracking challenge: Database collecting 2D and 3D time-lapse videos of moving cells from past and ongoing ISBI challenges. http://celltrackingchallenge.net/
- LiTS: Liver CT scans for tumor detection. https://competitions.codalab.org/competitions/17094
- LIDC-IDRI: Lung CT scans for cancer detection.
  https://wiki.cancerimagingarchive.net/display/Public/LIDC-IDRI
- DRIVE: A popular retinal fundus image dataset. https://drive.grand-challenge.org/

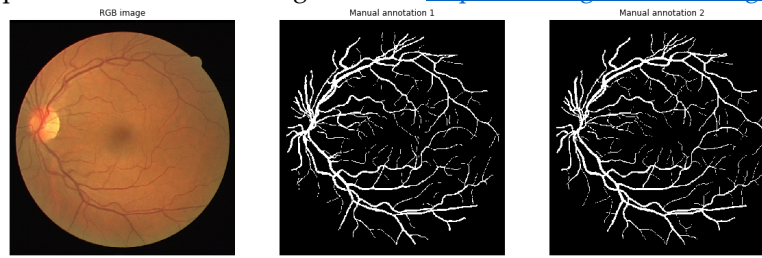

**Figure 13.** Fundus sample image along with corresponding ground truths from DRIVE [272].

- CT Colonography: CT scan dataset for colon cancer detection.
  https://wiki.cancerimagingarchive.net/display/Public/CT+COLONOGRAPHY
- Kaggle Data Science Bowl 2018: Nuclei segmentation challenge in microscopy images.
  https://www.kaggle.com/c/data-science-bowl-2018

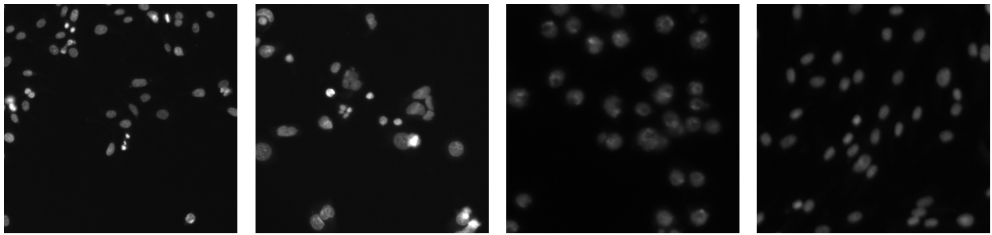

**Figure 14.** Example images from Kaggle data science bowl 2018 [273].

- ISIC archive: Database of Dermoscopy images from past and ongoing ISIC challenges. https://www.isic-archive.com/

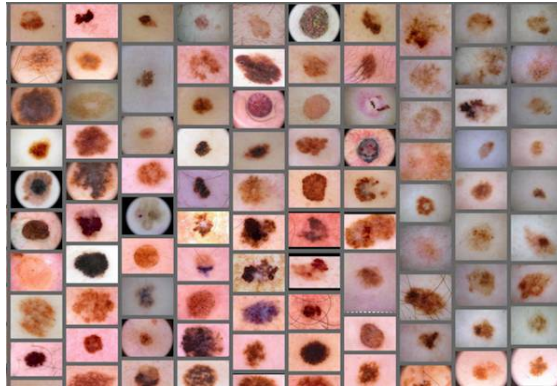

**Figure 15.** Examples of Dermoscopy images from the ISIC 2018 challenge [274], [275].

- SICAS Medical Image Repository: Archive for MICCAI Brain Tumor Segmentation Challenge (BRATS), MICCAI Ischemic Stroke Lesion Segmentation Challenge (ISLES), and ISBI Statistical Shape Model Challenge (SHAPE). https://www.smir.ch/Home/Browse

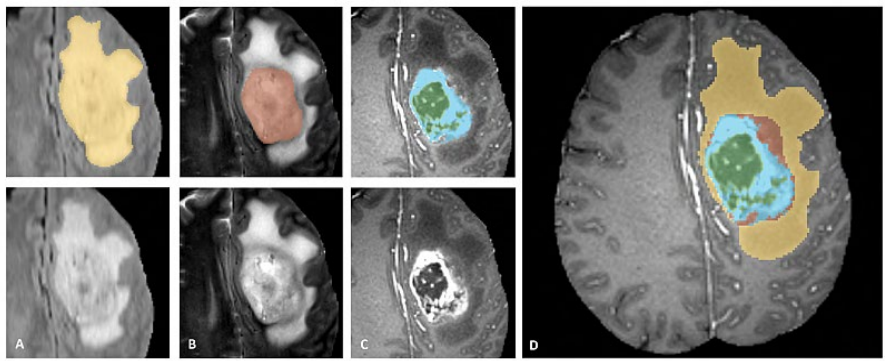

**Figure 16.** Example from Brain Tumor Segmentation Challenge (BRATS) 2017 [276]–[278].

- Medical Segmentation Decathlon: Collection of MR and CT databases for various target areas. http://medicaldecathlon.com/
- OASIS: Brain MRI and PET images. https://www.oasis-brains.org/
- ABIDE: Brain MRI datasets. http://fcon_1000.projects.nitrc.org/indi/abide/
- ICCVB: Prostate MRI and retinal fundus datasets. http://i2cvb.github.io/
- STARE: Retinal fundus dataset. http://cecas.clemson.edu/~ahoover/stare/

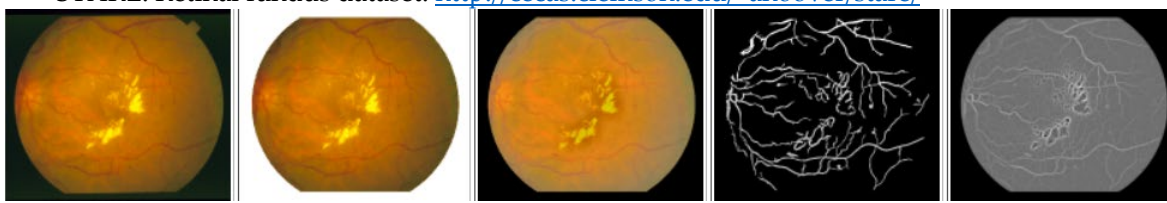

**Figure 17.** Example of STARE dataset showing the raw image, masked image, equalized image,

vessels, and MSF vessels [279], [280].

- CHASE_DB1: Retinal fundus dataset. https://blogs.kingston.ac.uk/retinal/chasedb1/
- SCR: Chest X-ray dataset. http://www.isi.uu.nl/Research/Databases/SCR/

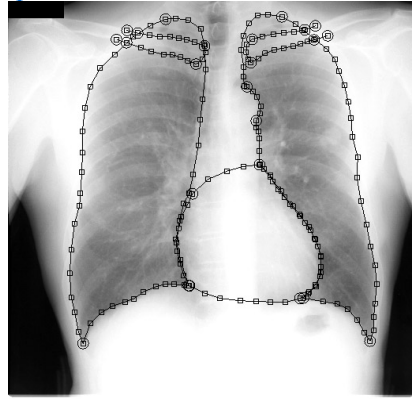

**Figure 18.** Example image from the SCR dataset. The lung field, the heart, and the clavicle are outlined [281], [282].

- DDSM: Mammogram dataset. http://www.eng.usf.edu/cvprg/Mammography/Database.html

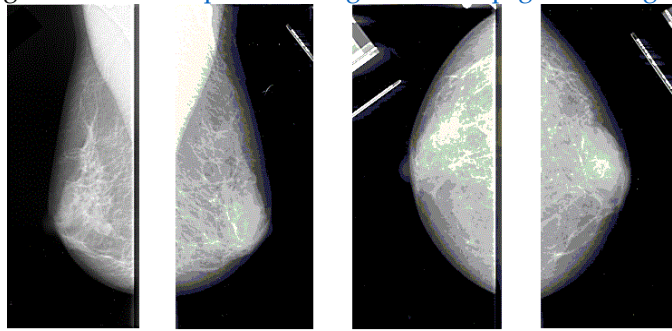

**Figure 19.** Sample images from the DDSM dataset [283], [284].

- BCDR: Mammogram database. https://bcdr.eu/
- mini-MIAS: Mammogram dataset. http://peipa.essex.ac.uk/info/mias.html
- PanNuke: Histology dataset for nuclei instance segmentation. https://jgamper.github.io/PanNukeDataset/

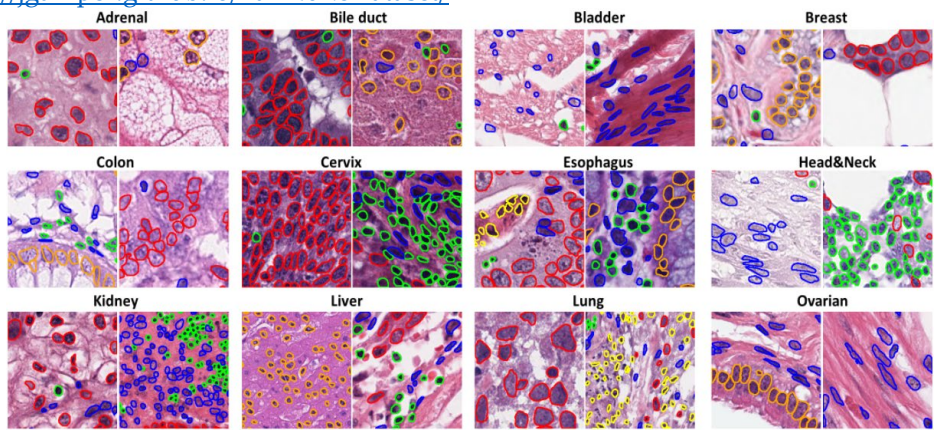

**Figure 20.** Ground truth of samples of different tissues from the PanNuke dataset [285], [286].

- University of Cyprus: Multiple sclerosis MRI, tele-orthopedics X-ray, and carotid ultrasound datasets. http://www.ehealthlab.cs.ucy.ac.cy/index.php/facilities/32-software/218-datasets
- The cancer imaging archive: A large public repository of cancer image datasets. https://www.cancerimagingarchive.net/

- Cardiac atlas project: Repository of cardiovascular image datasets. http://www.cardiacatlas.org/

*7.4 COVID-19 Datasets*

The following are some publicly available COVID-19 image datasets.

- COVID-CT: https://github.com/UCSD-AI4H/COVID-CT
- COVID-19 CT: http://medicalsegmentation.com/covid19/
- University of Montreal COVID-19 Image Data Collection: https://github.com/ieee8023/covid-chestxray-dataset
- RadiologyAi Consortium: https://www.radiologyaiconsortium.org/view

*7.5 Conferences & Journals*

Conferences:

- AAAI Conference on Artificial Intelligence (AAAI)
- British Machine Vision Conference (BMVC)
- Conference on Computer Vision and Pattern Recognition (CVPR)
- European Conference on Computer Vision (ECCV)
- International Conference on Computer Vision (ICCV)
- International Conference on Image Processing (ICIP)
- International Conference on Intelligent Robots and Systems (IROS)
- International Conference on Machine Learning (ICML)
- Medical Image Computing and Computer Assisted Intervention (MICCAI)
- Neural Information Processing Systems (NIPS)

Journals:

- IEEE Transactions on Image Processing
- IEEE Transactions on Medical Imaging
- IEEE Transactions on Pattern Analysis and Machine Intelligence
- International Journal of Computer Vision
- Journal of the American Medical Informatics Association
- Medical Image Analysis